\documentclass[10pt,conference]{IEEEtran}
\IEEEoverridecommandlockouts
\usepackage{cite}
\usepackage{amsmath,amssymb,amsfonts}
\usepackage{algorithmic}
\usepackage{textcomp}
\usepackage{hyperref}
\usepackage{graphicx}
\usepackage{booktabs}
\usepackage{makecell}
\usepackage{tablefootnote}
\usepackage[utf8]{inputenc}
\usepackage[table,xcdraw]{xcolor}
\def\BibTeX{{\rm B\kern-.05em{\sc i\kern-.025em b}\kern-.08em
    T\kern-.1667em\lower.7ex\hbox{E}\kern-.125emX}}
    
\begin{document}

\title{WALL*: A Web Application for Automated Quality Assurance using Large Language Models \thanks{WALL (\textbf{W}eb \textbf{A}pplication using \textbf{L}arge \textbf{L}anguage Models) is the name of our web application.}}

\author{\IEEEauthorblockN{ Seyed Moein Abtahi}
\IEEEauthorblockA{
\textit{Faculty of Engineering and Applied Science}\\
\textit{Ontario Tech University}\\
seyedmoein.abtahi@ontariotechu.net}
\and
\IEEEauthorblockN{ Akramul Azim}
\IEEEauthorblockA{
\textit{Faculty of Engineering and Applied Science}\\
\textit{Ontario Tech University}\\
akramul.azim@ontariotechu.ca}
}

\maketitle

\begin{abstract}
As software projects become increasingly complex, the volume and variety of issues in code files have grown substantially. Addressing this challenge requires efficient issue detection, resolution, and evaluation tools. This paper presents WALL, a web application that integrates SonarQube and large language models (LLMs) such as GPT-3.5 Turbo and GPT-4o to automate these tasks. WALL comprises three modules: an issue extraction tool, code issues reviser, and code comparison tool. Together, they enable a seamless pipeline for detecting software issues, generating automated code revisions, and evaluating the accuracy of revisions. Our experiments, conducted on 563 files with over 7,599 issues, demonstrate WALL’s effectiveness in reducing human effort while maintaining high-quality revisions. Results show that employing a hybrid approach of cost-effective and advanced LLMs can significantly lower costs and improve revision rates. Future work aims to enhance WALL's capabilities by integrating open-source LLMs and eliminating human intervention, paving the way for fully automated code quality management.
\end{abstract}


\section{Introduction}

As modern programming projects grow increasingly complex, the number of code files and programming languages used within a single project expands rapidly. Consequently, the variety and volume of issues that can arise in these files also increase significantly. Manually reviewing and debugging all files to ensure they are issue-free is a time-consuming and labor-intensive process. Even when a dedicated quality assurance team is in place, human error remains a significant risk in detecting and resolving all issues, especially in large projects. Leveraging automated frameworks for issue detection is highly beneficial to address this challenge. Tools like SonarQube conduct static code analysis to detect various software issues. These insights improve code quality, maintainability, and security. Software issues can result in system breakdowns, financial setbacks, and security risks, underscoring the importance of effective detection mechanisms \cite{becker2023programming}. While traditional detection methods offer some effectiveness, they often struggle to capture the nuanced and intricate problems characteristic of contemporary software development. To ensure code quality, developers commonly rely on static analysis tools \cite{yu2024security}.

Numerous studies have demonstrated the efficacy of static analysis tools in uncovering vulnerabilities that might be overlooked during manual code reviews \cite{9756950,ni2024learning}. Once issues are identified, the next critical step is their resolution. However, manually revising code to address all detected issues is both time-consuming and costly. Furthermore, humans may not consistently resolve all issues accurately. This research investigates the incorporation of large language models (LLMs) into issue detection frameworks, focusing on methodologies, challenges, and opportunities for enhancing code integrity assurance. LLMs can play a transformative role by automating the code generation process \cite{clark2024quantitative}. Utilizing the detailed issue reports generated by continuous code quality inspection platforms, LLMs significantly reduce the time and cost associated with manual corrections, thereby improving overall project efficiency.

The final step involves evaluating the revised code files generated by LLMs against the original files. Predefined metrics are employed to assess the quality of the revisions, ensuring that only files with unsatisfactory metrics require further inspection and review by the quality assurance team. When implemented within an automated pipeline, this streamlined approach ensures continuous code quality improvement with minimal human intervention.

\section{Related Work}

The adoption of large language models (LLMs) has garnered significant attention in software engineering research due to their potential to transform various stages of the software development lifecycle. Notably, Feng and Dong \cite{lin2024llm} examined the application of LLMs in code generation, illustrating their impact across multiple phases of software development. Similarly, Malik and Muhammad \cite{sami2024experimenting} explored the utility of a multi-agent paradigm powered by LLMs.

In addition to their role in automating development processes, LLMs have proven effective in identifying security vulnerabilities within software systems. Zhang et al. \cite{zhang2024prompt} highlighted how LLMs are increasingly being employed to detect software vulnerabilities.

The performance of LLM-based systems, however, is heavily influenced by the design of prompts. Yi and Gelei \cite{liu2024hitchhiker} emphasized the critical importance of prompt engineering, detailing how advanced prompt design and optimization techniques significantly enhance the effectiveness and reliability of LLM-driven solutions in achieving superior outcomes in software-related tasks. To further enhance the capabilities of LLMs, the Retrieval-Augmented Generation (RAG) technique has emerged as a powerful approach. This method integrates external knowledge sources with the model’s training data, resulting in improved performance in specialized tasks. Du et al. \cite{du2024vul} demonstrated the effectiveness of RAG in the domain of vulnerability detection, underscoring its potential to address complex security challenges in software systems.

Expanding on these advancements, Ali et al. \cite{ali2024establishing} proposed a novel framework to address requirements traceability, a critical aspect of software development and maintenance. Their approach leverages LLMs and RAG techniques to bridge the semantic gap between natural language requirements and software artifacts, such as UML class diagrams. By integrating keyword, vector, and graph-based indexing techniques, the method significantly enhances the accuracy and efficiency of traceability processes. Building on these studies, we developed the WALL tool to address the advancements and challenges highlighted in their research.

\section{Sources and Data}
\begin{figure}
    \centering
    \includegraphics[width=0.9\linewidth]{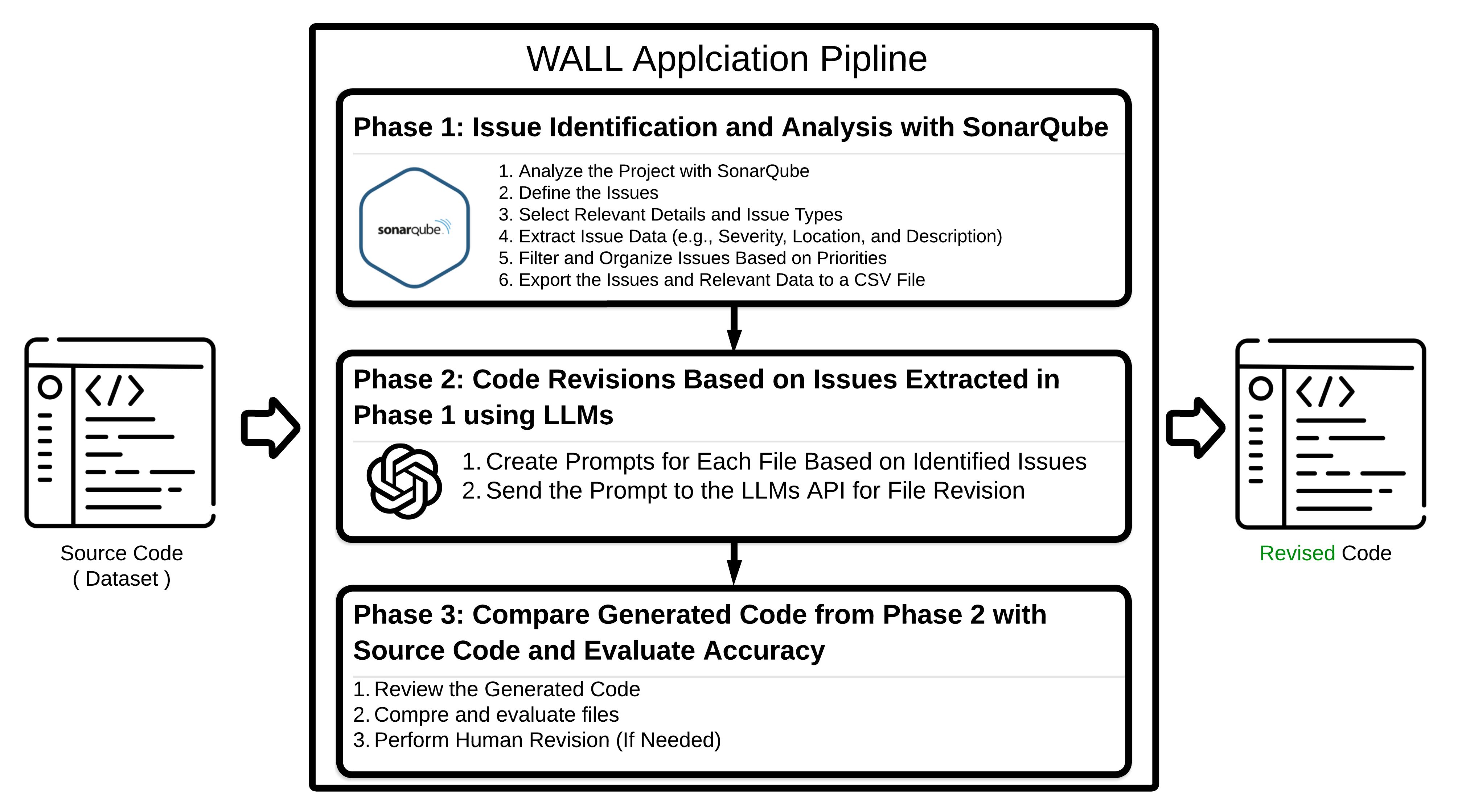}
    \caption{Overview of the WALL application workflow, including data processing steps and how the application functions}
    \label{fig:WALL_workflowl}
\end{figure}

This section outlines the sources and platforms employed in the design and development of the WALL tool and the datasets used for testing the application. The dataset comprises code files written in various programming and configuration languages. WALL comprises three core components: an Issue Extraction Platform, Code Revision using LLMs, and a Self-Developed Code Comparison Tool. The development and operation of the application follow these three sequential steps, ensuring a comprehensive and efficient workflow for detecting, revising, and evaluating code issues. An overview of the application workflow is illustrated in Fig. \ref{fig:WALL_workflowl}.

\subsection{Issue Extraction Platform}
\label{Issue Extraction Platform}
The static code analysis platform integrated into WALL is SonarQube \cite{sonarqube}, selected for its robust support of numerous programming languages, which is particularly advantageous for large software projects where maintaining high code quality across diverse languages is essential. SonarQube helps to identify bugs, which are unintentional errors made during development; code smells, which signal potential design or structural flaws that could benefit from refactoring; and vulnerabilities, which are weaknesses in the code that could pose security risks or be exploited \cite{sonarqube}.To achieve optimal results, it is recommended to use the latest version of SonarQube, which enhances issue detection capabilities and extends support for a wider range of programming languages.

\subsection{Large Language Models}
The language models selected for this experiment are GPT-3.5 Turbo \cite{gpt3}, GPT-4o, and GPT-4o Mini \cite{gpt4}. GPT-3.5 Turbo, a cost-effective and versatile model from OpenAI, is well-suited for general-purpose tasks. In contrast, GPT-4o, the latest and most advanced model, provides enhanced capabilities and improved performance over its predecessors. GPT-4o Mini offers a balance between performance and resource efficiency. The selection of these models follows a strategic approach: GPT-3.5 Turbo is used to revise the majority of files cost-efficiently, while GPT-4o is reserved for addressing complex issues that demand more sophisticated reasoning and access to up-to-date information.

\subsection{Code Comparison and Evaluation}

The WALL application features a side-by-side comparison of original and revised files, allowing users to identify differences between the two efficiently. This feature streamlines the evaluation process by highlighting changes, enabling more accurate assessments of revision quality and the feasibility of using LLMs for code modifications. Additionally, the tool detects code hallucinations or incorrect modifications generated by LLMs by marking inconsistencies, thus facilitating quicker edits and refinements as the final step in the workflow.

To quantitatively evaluate the accuracy of the revised files, WALL uses metrics such as precision, recall, and F1-score. These metrics, calculated based on the number of changed, original, and edited lines, provide a rigorous measure of revision quality \cite{hu2024unveiling}. While human intervention is currently required due to the absence of test cases for original files, future versions of WALL will eliminate this step. By using pre-existing test cases or those generated by LLMs, the tool will be able to automatically evaluate the revised files with test cases, paving the way for a fully automated pipeline to detect, revise, and validate code issues.

\subsection{Dataset}

For testing, we utilized proprietary project files provided by Team Eagle Ltd. \cite{team-eagle}, comprising over 2,500 files in languages such as Python, Java, JavaScript, and configuration and markup languages like Azure Resource Manager, CloudFormation, Docker, and HTML. Additionally, the open-source project "open-instruct" \cite{lambert2024tulu3} from GitHub repository \cite{open_instruct} was employed for a secondary experiment on a different dataset. The WALL application imposes no restrictions on the number of files or issues per file and supports all programming languages compatible with SonarQube \cite{sonarqube_lan}.

\begin{table*}[ht]
    \centering
    \caption{Example Format of Issues Extracted by SonarQube into CSV File}
    \resizebox{\textwidth}{!}{%
        \begin{tabular}{|c|c|c|c|c|}
            \hline
            \rowcolor[HTML]{67FD9A} 
            \multicolumn{1}{|c|}{\cellcolor[HTML]{67FD9A}File\_Location} & \multicolumn{1}{c|}{\cellcolor[HTML]{67FD9A}File\_Name} & \textbf{Line} & \multicolumn{1}{c|}{\cellcolor[HTML]{67FD9A}Message} & \textbf{Type} \\
            \hline
            \makecell{Project\textbackslash{}client\textbackslash{}src\\\textbackslash{}App.jsx} & App.jsx & 12 & A fragment with only one child is redundant. & CODE\_SMELL \\
            \hline
            \makecell{Project\textbackslash{}client\textbackslash{}src\\\textbackslash{}components\\\textbackslash{}OfflineControl.jsx} & OfflineControl.jsx & 116 & Visible, non-interactive elements with click handlers must have at least one keyboard listener. & BUG \\
            \hline
            \makecell{Project\textbackslash{}deploy\textbackslash{}helm\\\textbackslash{}dis\\\textbackslash{}deployment.yaml} & deployment.yaml & 20 & Specify a CPU limit for this container. & VULNERABILITY \\
            \hline
        \end{tabular}
    }
    \label{fig:code_issues}
\end{table*}

\section{WALL Web Application}

In this section, we present the design and backbone of the WALL tool and provide instructions for its usage. The source files for WALL are publicly available in a GitHub repository\footnote{\href{https://github.com/MoeinAbtahi/WALL}{https://github.com/MoeinAbtahi/WALL}}. The WALL Code Analysis Web Application is organized into three sequential modules, each of which must be executed to ensure a systematic and efficient workflow. The first module, the Issue Extraction Tool, identifies code issues through static analysis. The second module, the Code Revision Module, addresses these issues using OpenAI GPT models. Finally, the Code Comparison and Evaluation Module assesses the quality of the revisions. 

\subsection{Issue Extraction Tool}
\label{Issue Extraction Tool}
In this module, users must first add their project to SonarQube, either on a local machine or in the cloud, and obtain the project key and server URL to connect with SonarQube. This connection enables the extraction of issues from the desired project. The information required for the next phase (\ref{Code Issues Reviser Module}) includes the issue type, the line number where the issue occurs, the file location within the project, the file name, and the issue description. To facilitate this process, users are required to provide the SonarQube server URL, API token, and project key. Once these details are entered, the tool allows for extracting all necessary information into a single CSV (Comma-Separated Value) file, which serves as the input for the subsequent module, the Code Issues Reviser. Table \ref{fig:code_issues} presents an example of the CSV file output.

\subsection{Code Issues Reviser Module}
\label{Code Issues Reviser Module}
Based on the generated CSV file from the \ref{Issue Extraction Tool} module, users have two options. The first option allows them to upload the CSV file into the application, which will revise all identified files using pre-configured prompts and LLM models. The processing time for generating the revised files depends on factors such as the number of files, the length of each file, and the number of issues detected. The second option lets users manually select and view individual files from the CSV, generating revised versions one at a time. 

The prompt used in WALL was developed through a structured prompt engineering process involving over 50 experiments to reach the final version. This predefined prompt incorporates the code file, issue details from the CSV, a few-shot learning approach, and programming language specifications. In the "Processing All Files" mode, the prompt is fixed and cannot be modified, ensuring that the comparison of different GPT models is not influenced by varying prompt engineering techniques. Extensive testing was conducted to refine the prompt for optimal performance, eliminating other influencing factors. Conversely, the "One-by-One File Processing" mode allows the prompt to be editable, enabling users to customize it to meet specific requirements.

\subsubsection{Processing All Files}
A Python script for revising all files at once is available in the WALL GitHub repository. This script takes the generated CSV file from the Issue Extraction Tool as input. This method produces the revised files while maintaining the original project’s folder and file structure. However, the root folder name is appended with ``.Revised," and each revised file name is prefixed with ``Revised." 

\subsubsection{One-by-One File Processing}
This module, implemented in the current version of WALL, allows users to upload a CSV file and select specific files with issues. Users can switch between files as needed. A prewritten prompt is displayed upon selecting a file containing all the necessary information to enable the GPT model to revise the code effectively. Users also have the option to edit the prompt to tailor it to their specific needs. After finalizing the prompt, users can choose between GPT-3.5 Turbo, GPT-4o, or GPT-4o Mini to process the prompt. The generated response will vary depending on the selected model. Users can save the revised file in the same format as the original, with ``Revised." prefixed to the file name.

\subsection{Code Comparison Tool}
\label{Code Comparison Tool}

The final module of WALL is the Code Comparison and Evaluation Tool. This module enables users to compare original and revised files side-by-side, with differences highlighted for clarity. Lines removed from the original code are highlighted in yellow, while newly added lines in the revised version are marked in green. The tool also displays the file names and locations for both the original and revised files, making it easier for users to access and modify them if necessary. 

Additionally, the tool employs evaluation metrics such as F1-score, precision, and recall, adapted to assess changes at the line level. These metrics are calculated based on the number of lines changed, removed, or added between the original and revised versions of the code. In this version of WALL, the ground truth is determined by the percentage of updated, removed, or changed lines relative to the original files. In the absence of test cases, this approach assists the human reviewer in evaluating the accuracy of the GPT models. A high metric value (close to 100\%) indicates that the revised file closely mirrors the original, suggesting that the code had relatively few issues.

\section{Results and Analysis}

As a preliminary step, various experiments were conducted to create the prompt, as described in \ref{Code Issues Reviser Module}. Additionally, pre-experiments were performed on 350 issues across three categories (bugs, vulnerabilities, and code smells) to ensure fairness and verify that GPT-4o can resolve any issues GPT-3.5 Turbo can resolve.

In Experiment 1, WALL was evaluated on 7,599 issues across 563 files provided by Team Eagle Ltd. \cite{team-eagle}. These proprietary files are protected under organizational confidentiality policies and are not publicly accessible. The issues were categorized into code smells, bugs, and vulnerabilities. Using GPT-3.5 Turbo, WALL revised 5,441 issues, achieving a 71.6\% revision rate. With GPT-4o, the number of revised issues increased to 6,495, resulting in an 85.5\% revision rate. A detailed performance comparison is presented in Table \ref{table:gpt-performance-comparison}.

The tables show the success rate, defined as the percentage of issues resolved by each GPT model, and cost metrics derived from OpenAI's API usage data. The results demonstrate WALL's ability to reduce the time and effort required for code revision while maintaining cost efficiency. Its effectiveness in addressing a large volume of issues highlights its potential to enhance software quality improvement processes.
WALL was also tested on the open-source "open-instruct" repository \cite{open_instruct} as a secondary dataset. The results for this dataset are detailed in Table \ref{table:gpt-performance-comparison2}.

An optimized workflow to reduce costs involves a two-step process: first, revising files with GPT-3.5 Turbo, followed by a project rescan to identify unresolved issues, which are then addressed using GPT-4o. This hybrid approach can lower overall costs by up to 40\%, making it suitable for large-scale projects.

\begin{table}[ht]
\centering
\caption{Performance and Cost Comparison of GPT-3.5 Turbo and GPT-4o in Code Issue Revision in Experiment 1}
\resizebox{\columnwidth}{!}{
\begin{tabular}{@{}lcccc@{}}
\toprule
\textbf{Metric \textbackslash Type} & \textbf{Bugs} & \textbf{Vulnerability} & \textbf{Code Smell} \\ \midrule
\# Issues & \textbf{234} & \textbf{61} & \textbf{7304} \\ \midrule
\begin{tabular}[c]{@{}l@{}}GPT-3.5 Turbo Only\\ (\# Revised / Cost)\end{tabular} & 117 / \$1.68 & 59 / \$0.25 & 3,718 / \$8.13 \\ \midrule
\begin{tabular}[c]{@{}l@{}}GPT-4o for Remaining\tablefootnote{Remaining issues after the first revision with GPT-3.5 Turbo}\\ (\# Issues / Cost)\end{tabular} & 117 / \$3.08 & 2 / \$0.13 & 2,219 / \$18.69 \\ \midrule
\begin{tabular}[c]{@{}l@{}}GPT-4o Only\\ (\# Revised / Cost)\end{tabular} & \textbf{234} / \$6.20 & \textbf{61} / \$1.01 & 5,937 / \$32.57 \\ \midrule
\begin{tabular}[c]{@{}l@{}}GPT-3.5 + GPT-4o\\ (\# Revised / Cost)\end{tabular} & 234 / \$4.76 & 61 / \$0.38 & 5,937 / \$26.82 \\ \midrule
\begin{tabular}[c]{@{}l@{}}GPT-3.5 Success Rate\\ (\# Revised / Total)\end{tabular} & 50\% (117/234) & 96.7\% (59/61) & 50.9\% (3,718/7,304) \\ \midrule
\begin{tabular}[c]{@{}l@{}}GPT-4o Success Rate\\ (\# Revised / Total)\end{tabular} & \textbf{100\%} (234/234) & \textbf{100\%} (61/61) & 81.2\% (5,937/7,304) \\ \bottomrule
\end{tabular}
}
\label{table:gpt-performance-comparison}
\end{table}

\begin{table}[ht]
\centering
\caption{Performance and Cost Comparison of GPT-3.5 Turbo and GPT-4o in Code Issue Revision in Experiment 2 }
\resizebox{\columnwidth}{!}{
\begin{tabular}{@{}lcccc@{}}
\toprule
\textbf{Metric \textbackslash Type} & \textbf{Bugs} & \textbf{Vulnerability} & \textbf{Code Smell} \\ \midrule
\# Issues & \textbf{21} & \textbf{2} & \textbf{401} \\ \midrule
\begin{tabular}[c]{@{}l@{}}GPT-3.5 Turbo Only\\ (\# Revised / Cost)\end{tabular} & 19 / \$0.31 & 2 / \$0.01 & 369 / \$1.05 \\ \midrule
\begin{tabular}[c]{@{}l@{}}GPT-4o for Remaining\\ (\# Issues / Cost)\end{tabular} & 2 / \$0.59 & 0 / \$0.00 &  28 / \$0.66 \\ \midrule
\begin{tabular}[c]{@{}l@{}}GPT-4o Only\\ (\# Revised / Cost)\end{tabular} & 21 / \$2.19 & 2 / \$0.11 & 397 / \$8.35\\ \midrule
\begin{tabular}[c]{@{}l@{}}GPT-3.5 + GPT-4o\\ (\# Revised / Cost)\end{tabular} & 21 / \$0.9 & 2 / \$0.01 & 397 / \$1.71 \\ \midrule
\begin{tabular}[c]{@{}l@{}}GPT-3.5 Success Rate\\ (\# Revised / Total)\end{tabular} & \textbf{90.4}\% (19/21) & \textbf{100}\% (2/2) & \textbf{92}\% (369/401) \\ \midrule
\begin{tabular}[c]{@{}l@{}}GPT-4o Success Rate\\ (\# Revised / Total)\end{tabular} & \textbf{100\%} (21/21) & \textbf{100\%} (2/2) & \textbf{99}\% (397/401) \\ \bottomrule
\end{tabular}
}
\label{table:gpt-performance-comparison2}
\end{table}

\section{Limitations and Future Work}  
The WALL tool is an agile software project in its initial phase, with plans for updates and feature enhancements. One key goal is to integrate open-source LLMs, such as the LLAMA models \cite{touvron2023llama}\cite{llama2024}, for a more cost-effective solution to code revision tasks. Additionally, we plan to adopt the Retrieval-Augmented Generation (RAG) approach to broaden the tool's capabilities while leveraging affordable models. Future iterations will also explore fine-tuning techniques to improve revision accuracy and quality. As outlined in Section \ref{Code Comparison Tool}, a major objective is to automate the workflow, eliminating human intervention for a seamless pipeline that reduces time and cost in improving software quality for large-scale projects. Current limitations in the RAG module, like complex query handling and retrieval accuracy, will be addressed in future updates.

\section*{Conclusion}
This study highlights WALL as a transformative solution for enhancing software quality through automation. By leveraging SonarQube for issue detection and LLMs for automated code revisions, WALL significantly reduces the time and cost associated with manual debugging and code refinement. The tool's modular design allows for flexibility in issue resolution, from batch processing to fine-tuned revisions. Our results demonstrate high revision success rates, especially with a hybrid approach combining GPT-3.5 Turbo and GPT-4o, optimizing both performance and cost. WALL’s built-in evaluation metrics ensure the quality of revisions, marking a significant step toward automating software maintenance workflows. Future developments will focus on integrating more cost-effective LLMs and automating the entire pipeline, eliminating the need for human oversight. This evolution will further streamline software development processes, enhancing productivity and code reliability across diverse programming environments. 
\bibliography{Reference}
\bibliographystyle{IEEEtran}

\end{document}